# EFFICIENCY LIMITS OF QUANTUM WELL SOLAR CELLS

J.P.Connolly[1], I.M.Ballard[2], K.W.J.Barnham[2], D.B.Bushnell[3], T.N.D.Tibbits[2], J.S.Roberts[4]
1. LECA (UMR 7575 CNRS), Ecole Nationale Supérieure de Chimie Paris, Paris 75005 (j.connolly@ic.ac.uk)
2. Blackett Laboratory, Physics Department, Imperial College London, London SW7 2BW, U.K.
3. International Rectifier Ltd , Hurst Green, Oxtead, Surrey, RH8 9BB.
4. EPSRC National Centre for III-V Epitaxy, University of Sheffield, Sheffield S1 3JD, U.K.

ABSTRACT: The quantum well solar cell (QWSC) [1] has been proposed as a flexible means to ensuring current matching for tandem cells [2]. This paper explores the further advantage afforded by the indication that QWSCs operate in the radiative limit because radiative contribution to the dark current is seen to dominate in experimental data at biases corresponding to operation under concentration. The dark currents of QWSCs are analysed in terms of a light and dark current model. The model calculates the spectral response (QE) from field bearing regions and charge neutral layers and from the quantum wells by calculating the confined densities of states and absorption coefficient, and solving transport equations analytically [3]. The total dark current is expressed as the sum of depletion layer and charge neutral radiative and non radiative currents consistent with parameter values extracted from QE fits to data. The depletion layer dark current is a sum of Shockley-Read-Hall non radiative [6], and radiative [4] contributions. The charge neutral region contribution is expressed in terms of the ideal Shockley radiative and non-radiative currents [5] modified to include surface recombination. This analysis shows that the QWSC is inherently subject to the fundamental radiative efficiency limit at high currents where the radiative dark current dominates, whereas good homojunction cells are well described by the ideal Shockley picture where the limit is determined by radiative and non radiative recombination in the charge neutral layers of the cell.

## 1 INTRODUCTION

The QWSC [1] is a p-i-n structure with quantum wells in the field bearing intrinsic region, sandwiched between higher gap barriers. It has been suggested that the increase in photocurrent is greater than the increase in dark current, leading to higher efficiency limits for this design than for bulk solar cells. Photogenerated carriers escape the wells with close to unit efficiency for applied bias up to the operating voltage [3] and contribute to the photocurrent.

In order to evaluate the potential efficiency limits of these cells, we model the photocurrent in terms of diffusion currents from charge neutral layers, and of highly efficient carrier collection for carriers photogenerated in the field bearing layers, in close agreement with experiment [3,15]. We calculate the dark current in terms of radiative and non radiative contributions.

In the limit where superposition holds, this yields the efficiency in terms of the sum of photocurrent and dark current.

The study includes strained structures where the use of alternating barrier and well layers is exploited to balance opposite stresses in a technique known as strain balancing [9] which allows many periods of strained layers to be grown without the formation of dislocations.

## 2 MODEL

### 2.1 Photocurrent model

The FORTRAN QWSC spectral response model SOL has been described previously [3,15]. Briefly, it consists of solving transport equations for p and n layers subject to surface recombination and depletion approximation boundary conditions. The carriers photogenerated in the space charge region are assumed to be collected with unit efficiency, in good agreement with measurement.

It has been extended to apply to a range of materials and to incorporate the effects of strain [8]. Absorption coefficients are unavailable for strained materials, which do not exist in bulk form. They are estimated by interpolating neighbouring absorption data sets over energy and wavelength to account for compositional and strain bandgap shift effects.

Absorption coefficients for a continuous range of lattice matched compositions in the InGa(y)As(x)P on GaAs and InP substrates, and Al(x)GaAs on GaAs substrates, are interpolated between available end-points from data in the literature [13,14].

To calculate an absorption coefficient for a strained layer, the closest lattice matched absorption coefficient is found and shifted by the difference in bandgap identified from the difference in lattice constant and resulting strain, and the elastic constants for the material as defined in published data [10,12,13]. Although the additional parameters required necessarily increase the uncertainty in quantum well absorption coefficients in particular, we will see that this approach is sufficiently accurate to date without taking effects such as effective mass strain dispersion into account.

Variable growth dependent parameters in the QE data fits are exciton strength and broadening, minority carrier diffusion lengths in charge neutral layers, and surface recombination velocities. All other parameters are derived from the literature.

The integral of the QE and a given spectrum then yields the short circuit current $I_{SC}$.

### 2.2 Charge neutral layer ideal Shockley dark current

The ideal Shockley dark current model [5] is derived by considering majority carrier diffusion as a function of bias across the space-charge region. Accordingly, the resulting recombination current includes both radiative and non radiative contributions. It is expressed as

$$J_S = q \left( e^{\frac{qV}{K_B T}} - 1 \right) \left( \frac{q.ni_p^2}{N_D} \frac{D_n}{L_n} + \frac{q.ni_n^2}{N_D} \frac{D_p}{L_p} + S_N \frac{q.ni_p^2}{N_A} e^{-x_p/L_n} + S_P \frac{q.ni_n^2}{N_D} e^{-x_n/L_p} \right) \quad (1)$$

for applied bias V, and charge neutral p and n widths $x_p$ and $x_n$. The other variables have their usual definitions. Equation (1) is the ideal Shockley dark current with the addition of surface recombination terms which are included for completeness but are negligible in this study.

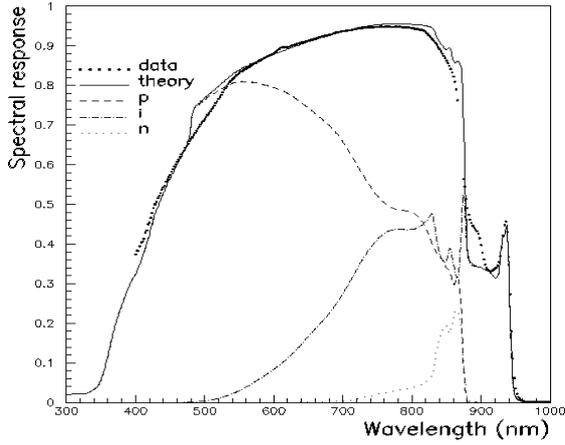

**Figure 1** Quantum efficiency data and theory for 50 well QT1744

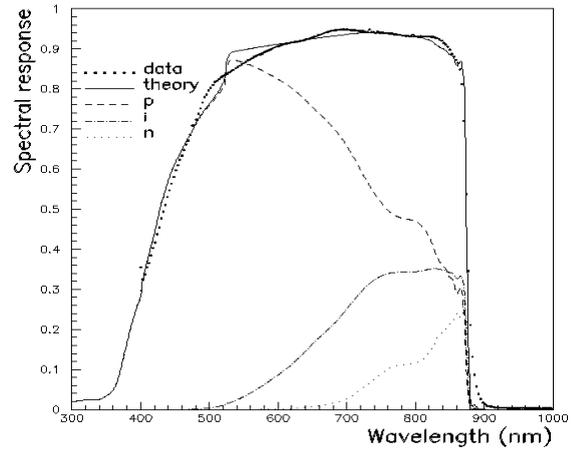

**Figure 2** GaAs p-i-n spectral response

The diffusion lengths $L_n$ and $L_p$ and recombination velocities $S_N$ and $S_P$ are obtained by adjusting guidelines suggested by SOL from the literature in order to fit experimental QE. This is possible because the diffusion length affects the QE across the wavelength range, whereas recombination velocity losses are stronger at shorter wavelengths because of increasing absorption coefficient.

2.3 SCR non radiative and radiative dark currents

The non radiative recombination rate U via mid-gap trapping centres at position $x$ in the space charge region is described in the Shockley-Read-Hall methodology as

$$U_{SRH}(x) = \frac{pn - n_i^2}{\tau_n.(p + p_{trap}) + \tau_p.(n + n_{trap})} \quad - (2)$$

where carrier densities n(x) and p(x) are calculated as in ref. [11], and may vary according to local potential at x and whether x is in a barrier or a quantum well layer. The trap densities $p_t$ and $n_t$ are calculated for the dominant mid-gap traps.

The non-radiative depletion layer dark current is then the integral of U(x) over intrinsic layer width $x_i$ and p and n depletion widths $x_{wp}$ and $x_{wn}$

$$J_{SRH} = \int_0^{x_{wp}+xi+x_{wn}} U(V,x).dx \quad - (3)$$

In lattice matched materials the lifetimes $\tau_n$ and $\tau_p$ can be estimated by fitting homogeneous control p-i-n structures [7]. For strained material this is not possible, and $\tau_n$ and $\tau_p$ are assumed to be described by a single value which represents an lifetime averaged by the carrier densities and recombination rates.

As a result the expression for $J_{SRH}$ is fitted with a single lifetime parameter $\tau$.

The depletion layer radiative recombination current $J_{RAD}$ is obtained [4] from the generalised Planck equation for a non-black body. This relates (equation 4) absorptivity $\alpha$ to emission for electron-hole populations with quasi-Fermi levels separated by potential $q\phi$. The double integral is over position x which as in equation (3) ranges over depleted widths $x_{wp}$, $x_{wn}$ and $x_i$, and an angular integral over front and back surfaces S

$$J_{RAD} = \int^{x_{wp}+x_i+x_{wn}} \int^S \frac{2n^2}{h^3 c^2} \frac{E^2 \alpha(E)}{e^{(E-q\phi)/kT}-1}.dS.dx \quad - (4)$$

The quasi-Fermi level separation $q\phi$ may be position dependent, but is assumed constant and equal to the applied bias here. Points to note are that the absorptivity $\alpha(E)$ is known from the spectral response calculation and data fits, and $J_{RAD}$ is the total emission, not that emitted through the front surface.

2.4 Efficiency

The efficiency of the solar cell is then found assuming superposition from the light current voltage (IV) characteristic:

$$I(V) = I_{SC} - (J_S + J_{SRH} + J_{RAD}) - (5)$$

| QT1774A (QWSC) |
|---|
| Window p+-$Al_{0.8}$GaAs    2E24 $m^{-3}$ |
| Emitter p-GaAs             2E24 $m^{-3}$ |
| i region 50 periods of: <br> {98Å barrier $GaAsP_{0.089}$ + 85Å well $In_{0.12}GaAs$} <br> Total width 1.45μm |
| Base n-GaAs               2E24 $m^{-3}$ |
| QT510c (p-i-n) |
| Window p-$Al_{0.7}$GaAs        2.3E24 $m^{-3}$ |
| Emitter p-GaAs 0.5μm         2.3E24 $m^{-3}$ |
| Intrinsic GaAs 0.9μm    ~1E21 $m^{-3}$ p-type <br>                          Background doping |
| Base n-GaAs 2μm              1.8E25 $m^{-3}$ |

**Table I**: Structures of QWSC qt1744 and p-i-n qt510c

2   EXPERIMENTAL DETAILS

Samples were grown by MOVPE and processed into 1mm diameter circular mesa structures with total area $7.85 \cdot 10^{-7}$ $m^2$ and active area $2.33 \cdot 10^{-7}$ $m^2$. To allow study of the radiative currents mesas were fully metallised to reduce series resistance effects masking high current dark current behaviour. This does not significantly change the radiative dark current calculation since the luminescence is largely emitted into the substrate, and the escaping fraction is negligible. The difference is nevertheless taken into account by setting

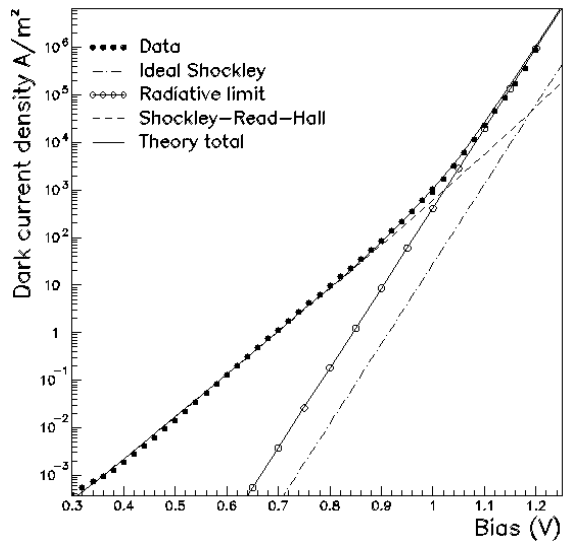

**Figure 3** Dark current fit for 50 well QT1744 showing radiative dominated behaviour above ~1V

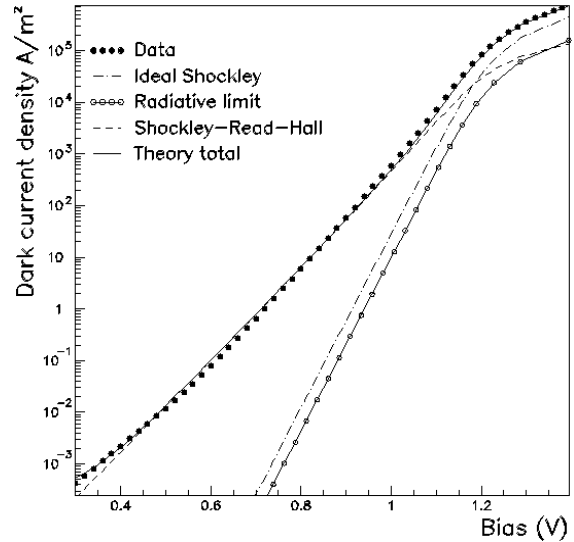

**Figure 4** Dark current for p-i-n showing ideal Shockley dominated behaviour past ~1.1V

the active area to zero in dark current calculations for fully metallised samples.

Spectral response measurements were carried out using a computer controlled Bentham monochromator and lockin amplifier.

Efficiencies were measured under a tungsten-halogen light source calibrated to provide illumination levels equivalent to the AM1.5 global (ASTM standard) spectrum determined by a calibrated GaAs cell.

Table 1 shows sample details for the QWSC and p-i-n structures as defined in the modelling.

4      MODELLING

4.1) Quantum efficiency

Figure 1 shows quantum efficiency for QWSC 1744. The response in the region of the QW to bulk transition at about 880nm is slightly overestimated but the exciton fits closely. The surface recombination velocity is high, indicating poor interface quality at the interface between the GaAs emitter and the Al(82)GaAs window.

The fit fixes bulk transport parameters in the charge neutral regions, and quantum well absorption parameters which are exciton strength and broadening, which respectively determine ideal Shockley and radiative recombination levels.

The GaAs p-i-n fit is shown in figure 2 for comparison.

4.2) Dark current

Figure 3 shows the corresponding QWSC dark current fit. A single lifetime of 15ns fits the SRH dark current which dominates until 1V. Above 1V the radiative contribution from the intrinsic region dominates, with the ideal Shockley current about an order of magnitude lower.

Figure 4 shows the same fit for the p-i-n cell which is not fully metallised and shows series resistance effects, and parallel resistance at low bias.

The total dark current (solid line) is the sum of $J_{SRH}$ (dashed) dominating at low bias, $J_{RAD}$ space-charge region radiative (solid ringed) and ideal Shockley $J_S$ and shows a very close fit in both cases.

The transition from ideality 2 to ideality 1 behaviour is standard, but this analysis indicates that the QWSC pictured here is radiatively dominated at bias above a volt, significantly below the built in voltage of 1.399V in this case.

The same may be true of the GaAs cell but cannot be stated with certainty because in this case the transition from slope 2 to slope 1 corresponds to transition to the ideal Shockley which includes both radiative and non radiative contributions. The non radiative lifetime in this case is 10ns.

The n=1 regime in the p-i-n is therefore only fully radiative in the limit of an ideal cell, whereas the QWSC is radiatively limited even in the case of non ideal bulk recombination.

It is worth noting that the radiative current cannot be adjusted since it lacks any free parameters. All factors in the expression for $J_{SRH}$ are determined by device design (geometry) and by QE fitting (optical functions).

The lower bias SRH dominated range is subject to more uncertainty because of the lack of information concerning carrier lifetimes, and the necessity of reducing the lifetime parameters to a single lifetime.

In the QWSC case, the lifetime is of 15ns which is in the range of good lifetimes for intrinsic layers in high quality GaAs.

Finally it can be seen that the QWSC brings the transition from ideality 2 to ideality 1 behaviour to lower bias because of the increased importance of radiative recombination which is accompanied by increased absorption.

This analysis therefore suggests that the QWSC is favourable because QWSCs are intrinsically dominated by radiative recombination and therefore are subject to the higher efficiency limits governing solar cells.

The same analysis has been applied to high efficiency published data [16] and shows identical behaviour with a similar SRH lifetime of 16ns.

4.3) Efficiency

The efficiency of cell qt1744 under spectrum AM1.5 global (ASTM standard) was measured at 21.3% active area efficiency, meaning without contact shading loss. The model predicts 21.0% from the theoretical QE and dark current, and 21.1% from experimental QE and measured dark current. The difference is due to the experimental light IV predicting a slightly higher (+3%) short circuit current which is within acceptable efficiency measurement uncertainty due to the light source.

These cells are designed for terrestrial concentration applications. In order to estimate efficiency limits in this picture we assume no shading and 100% absorption in the wells, which requires light trapping techniques in practise. For concentrations of 400, we obtain 34.0% efficiency for AM1.5 global and 33.0% for AM1.5 direct using superposition and a dark current as calculated in figure 3. We choose 400x because the operating voltage is then 1.1 – 1.14V, in the radiative regime.

We emphasize that this is a idealised treatment but therefore comparable with classic efficiency limits.

These are not limiting efficiencies in the final sense since they do not address the issue of position dependent quasi-Fermi-level separation $q\phi$ mentioned above. Previous electroluminoescence studies of single wells [4] has found evidence of $q\phi$ being lower in the well than in the barriers, for $q\phi$ is determined by the applied bias as in this paper.

This study of dark currents however lacks the sensitivity to explore this issue, since the effective shift of some tens of mV in the radiative dark current contribution is below the uncertainty in the measurement.

4       CONCLUSIONS

We have presented quantitative fits of the QE of strained QWSCs. We have described the close interaction between QE fits which determine transport and optical parameters, and a description of the QWSC and homostructure dark current in terms of charge neutral and space-charge region recombination currents. This includes non radiative and radiative contributions.

The dark current model has no free parameters in the radiative regime at high bias, because parameters are fixed by the QE fits. It fits QWSCs and p-i-ns closely.

This analysis shows that good p-i-n cells show a transition to ideal Shockley behaviour which however includes non radiative contributions.

We have shown that the QWSC demonstrates ideal radiative behaviour since the analysis shows that the dark current is dominated by radiative recombination

Further study of electroluminescence of these cells is required to answer the question of limiting efficiency for these structures, because the dark current analysis lacks the sensitivity to answer questions concerning quasi-Fermi level separation in the space charge region.

**Acknowledgements:**
Jenny Nelson's comments, clarifications, and advice are gratefully acknowledged, as is Paul Stavrinou's help with materials parameters and models

**References:**
1. K.W.J. Barnham and G. Duggan, "A New Approach to Multi Bandgap Solar Cells", J. Appl. Phys., 67, pp. 3490 (1990)
2. J. P. Connolly, K. W. J. Barnham, J. Nelson, C. Roberts, M. Pate and J. S. Roberts. "Short Circuit Current Enhancements in GaAs/Al$_x$-Ga$_{1-x}$As MQW Solar Cells". 2nd World Conference and Exhibition on Photovoltaic Solar Energy Conversion, Vienna, 3631 (1998).
3. M.Paxman, J.Nelson, K.W.J.Barnham, B.Braun, J.P.Connolly, C.Button, J.S.Roberts and C.T.Foxon, "Modelling Modelling the Spectral Response of the Quantum Well Solar Cell" *J.Appl.Phys.* 74, 614 (1993).
4. J. Nelson, J. Barnes, N. Ekins-Daukes, B. Kluftinger, E. S-M Tsui, K. W. J. Barnham, C. T. Foxon and J. S. Roberts. "Observation of Suppressed Radiative Recombination in Single Quantum Well P-I-N Photodiodes. J. Appl. Phys., 82, 6240 (1997).
5. W. Shockley "Electrons and Holes in Semiconductors", D. Van Nostrand, Princeton, N.J., 1950
6. W.Shockley and W.T.Read, *Phys. Rev.* **87** (1952) 835, R.N.Hall, *Phys. Rev.* **87**, (1952) 387.
7. James P. Connolly, Jenny Nelson, Keith W.J. Barnham, Ian Ballard, C. Roberts, J.S. Roberts, C.T. Foxon, "Simulating Multiple Quantum Well Solar Cells", Proc. 28th IEEE Photovoltaic Specialists Conference, Anchorage, Alaska, USA, Sept. 2000, p. 1304-1307.
8. James P. Connolly, Jenny Nelson, Keith W.J. Barnham, Carsten Rohr, Chris Button, John Roberts, Tom Foxon, "Modelling Multi Quantum Well Solar Cell Efficiency", Proc. 17th European Photovoltaic Solar Energy Conference, Munich, Germany, Oct. 2001 (WIP, Munich and ETA, Florence, 2002) pp 204-207
9. N J Ekins-Daukes, J Zhang, D B Bushnell, K W J Barnham, M Mazzer, J S Roberts, "Strain-balanced Materials for High-efficiency Solar Cells", *Proc. 28th IEEE Photovoltaic Specialists Conference*, (IEEE, New York, 2000) p1273
10. Introduction to Solid State Physics, Charles Kittel, Wiley 1995
11. J.Nelson, I.Ballard, K.W.J.Barnham, J.P.Connolly et al. *J.Appl.Phys.*, **86**, (1999) 5898.
12. Paul Stavrinou, EXSS, Imperial College Londo, private communication
13. In: Properties of III-V Seciconductor Materisls, INSPEC EMIS Datareviews Series
14. SM Sze Physics of semiconductor Devices
15. J.Barnes, J.Nelson, K.W.J.Barnham, J.S.Roberts, M.A.Pate, S.S. Dosanjh, R.Grey, M.Masseur and F.Ghiraldo,"Characterisation of GaAs/InGaAs quantum wells using photocurrent spectroscopy", J. Appl. Phys. {\bf 79} 7775-9 (1996)
16. Sarah R. Kurtz,*et al.* "High efficiency GaAs solar cells using GaInP$_2$ window layers", 0160-8371/90/0000-0138 1190 IEEE